\documentclass[journal,article,submit,pdftex,moreauthors, computers]{Definitions/mdpi}

\usepackage{amsmath,amssymb,amsfonts}
\usepackage{algorithmic}
\usepackage{graphicx}
\graphicspath{{./figures/}}
\usepackage{textcomp}
\usepackage{xcolor}
\usepackage{tabularx}
\usepackage{url}
\usepackage{balance}
\usepackage{mathtools}
\usepackage{color}
\usepackage{booktabs}
\usepackage[linesnumbered,lined,boxed]{algorithm2e}
\usepackage{siunitx}
\SetAlCapSkip{1em}
\usepackage{subcaption}
\usepackage{nth}

\usepackage{pgfplots}
\pgfplotsset{compat=newest}
\usepgfplotslibrary{groupplots}
\usepgfplotslibrary{dateplot}

\firstpage{1} 
\pubvolume{1}
\issuenum{1}
\articlenumber{0}
\pubyear{2022}
\copyrightyear{2022}
\datereceived{} 
\dateaccepted{} 
\datepublished{} 
\hreflink{https://doi.org/} 

\Title{A Light Signaling Approach to Node Grouping for Massive MIMO IoT Networks}

\TitleCitation{A Light Signaling Approach to Node Grouping for Massive MIMO IoT Networks}

\Author{Emma~Fitzgerald$^{1}$, Micha{\l} Pi\'{o}ro$^{2}$, Harsh~Tataria$^{1}$, Gilles Callebaut$^{3}$, Sara Gunnarsson$^{1}$$^{3}$ and Liesbet Van der Perre$^{3}$$^{1}$
 \thanks{The work of
 E.~Fitzgerald was supported by the Celtic-Next project IMMINENCE, the SSF project SEC4FACTORY under grant no. SSF RIT17-0032, and the
 strategic research area ELLIIT. The work of M.~Pi\'{o}ro was supported by the National Science
Centre, Poland, under the grant no. 2017/25/B/ST7/02313: ``Packet
routing and transmission scheduling optimization in multi-hop
wireless networks with multicast traffic''. The work of H.~Tataria was partly supported by
 Ericsson AB, Sweden.}}

\AuthorNames{Emma~Fitzgerald, Micha{\l} Pi\'{o}ro, Harsh~Tataria, Gilles Callebaut, Sara Gunnarsson and Liesbet Van der Perre}

\AuthorCitation{Fitzgerald, E.; Pi\'{o}ro, M.; Tataria, H.; Callebaut, G.; Gunnarsson, S.; Van der Perre, L.}

\address{%
$^{1}$ \quad Department of Electrical and Information Technology, Lund University, SE-221 Lund, Sweden; \{emma.fitzgerald, harsh.tataria@eit.lth.se, sara.gunnarson\}@eit.lth.se\\
$^{2}$ \quad Institute of Telecommunication, Warsaw University of Technology, Nowowiejska 15/19, 00-665 Warsaw, Poland;  m.pioro@tele.pw.edu.pl\\
$^{3}$ \quad Department of Electrical Engineering, KU Leuven, Gebroeders de Smetstraat 1, 9000 Ghent, Belgium; \{gilles.callebaut, liesbet.vanderperre\}@kuleuven.be}

\corres{Correspondence: emma.fitzgerald@eit.lth.se}

\abstract{Massive MIMO is a leading technology to connect very large numbers of energy
constrained nodes, as it offers both extensive spatial multiplexing and large
array gain. A challenge resides in partitioning the many nodes into groups that
can communicate simultaneously such that the mutual interference is minimized.
We here propose node partitioning strategies that do not require full channel
state information, but rather are based on nodes' respective directional
channel properties. In our considered scenarios, these typically have a time
constant that is far larger than the coherence time of the channel. We developed
both an optimal and an approximation algorithm to partition users based on
directional channel properties, and evaluated them numerically. Our results show
that both algorithms, despite using only these directional channel properties, achieve similar performance in terms of the minimum signal-to-interference-plus-noise ratio for any user, compared with a reference
method using full channel knowledge. In particular, we demonstrate that grouping nodes with related directional properties is to be avoided. We hence
realise a simple partitioning method requiring minimal information to be
collected from the nodes, and where this information typically remains stable over a long term, thus promoting their autonomy and energy efficiency.}

\keyword{massive MIMO; IoT; user grouping; energy efficiency}

\begin{document}

\nolinenumbers

\section{Introduction}\label{sec:intro}
It is clear by now that massive multiple-input multiple-output (MIMO) is a key
technology for achieving high capacity in broadband wireless networks, and is currently being deployed to this end in expanding 5G networks.
The use of many electronically steerable service antennas at the base station
facilitates aggressive spatial multiplexing to tens of user equipment nodes.
This enables a substantial increase in the link reliability experienced by a particular node. In situations where the massive MIMO array
is required to simultaneously serve a very large number of nodes, there arises
the challenge of first grouping and then scheduling these nodes for
transmission. This is particularly important for the massive machine-type
communication (mMTC) use case, where extremely large numbers of nodes may be
present within the same geographical area served by the base station
\cite{3gpp2010further}. 

In this paper, we propose a node grouping method based on directional properties
of the nodes' channels that are stable over long periods of time and so do not
need to be updated at the rate of the coherence time of the channel. This
reduces signalling overhead and makes our method suitable for use as a
precursor to node scheduling for energy constrained Internet of Things (IoT)
nodes with infrequent transmissions.  We formulate this as an optimization
problem using mixed-integer programming. We then develop an efficient
approximation algorithm to solve this problem much faster than is possible using
the optimization process. Our results show that both our optimization and approximation
provide superior performance in terms of minimum
signal-to-interference-plus-noise ratio (SINR) in comparison to a reference
method based on grouping nodes by their instantaneous signal strength, that is,
received power taking into account both small and large-scale fading. The
reference method requires full small scale channel state information (CSI), however the dominant power consumption of IoT nodes is due to transmission, as shown in~\cite{s21030913}. Hence, sending pilot sequences, to learn small-scale CSI, prior to data, would consume too much energy. In contrast, less time-varying information could be reused without the need to relearn the channel for each uplink message.

The contributions of this paper are:
\begin{itemize}
    \item A node partitioning method for massive MIMO systems based on simple
    	directional characteristics of the nodes' channels, namely dominant direction and
	angular spectrum spread, allowing for stable partitioning over a long time
	period with minimal signalling by the node devices. The partitioning is
	complete in the sense that every node belongs to a group, so that all
	nodes are served.
    \item A formulation of the node partitioning problem using directional channel
    	characteristics as an optimization problem using mixed integer
	programming. 
    \item An efficient approximation algorithm to solve the problem that runs in
	$O(n\mathrm{log}n)$ time, where $n$ is the number of nodes. 
    \item A numerical evaluation of the optimization problem and approximation
    	algorithm using network examples generated with a suitable channel
    	model, showing similar performance for our partitioning method compared
    	to using full channel information.
\end{itemize}

The rest of this paper is organised as follows. Section~\ref{sec:motivation}
introduces and motivates our node grouping approach. Section~\ref{sec:related_work}
then discusses related work on node partitioning and scheduling in massive MIMO
systems. Next, Section~\ref{sec:scenario} details our problem setting and defines
the node partitioning problem. The partitioning problem is then formulated as a
mixed-integer optimization problem in Section~\ref{sec:system_model}, and our
approximation algorithm to solve this problem more efficiently is given in
Section~\ref{sec:approximation}. We present our numerical evaluation of the optimization
problem and approximation algorithm in
Section~\ref{sec:evaluation}. Finally, in Section~\ref{sec:conclusion}, we
conclude this paper and discuss future work.

\section{Motivation}\label{sec:motivation}

Many IoT devices are battery-powered and transmit only infrequently, spending
the rest of their time in a sleep state.\footnote{We
recognise that some IoT devices may need to constantly send or receive data,
however here we focus on typical sensors with infrequent transmissions.} Among
these devices, there is a large variation in the timescale when transmission
and reception need to take place, with the absolute timescale varying from
minutes up to tens of days or longer. At such time scales, efficient node
grouping and scheduling becomes difficult, since information about the node
traffic and propagation channels can quickly become outdated, unless additional
signaling is introduced, which in turn requires nodes to wake up and transmit
more often, consuming more energy.

Node grouping consists of partitioning the nodes into groups such that each
group can be accommodated for simultaneous transmission in the same coherence
block, giving a combination of spatial and time multiplexing. This subsequently
allows scheduling of node groups rather than individual nodes, thus reducing the
complexity of the scheduling problem. For low-power IoT devices, this partition
should be based on node properties that are stable over considerable time, so
that schedules can cover a reasonable number of transmissions, thus reducing
signaling and computation costs for scheduling.  In addition, node properties
used for grouping must also be informative for network performance. To this end,
interference within the node groups should be minimized by exploiting
characteristics of propagation channels that remain stable over a long time,
ideally at least over some tens of node transmissions, and as long as possible
to allow for long scheduling windows. This is the subject of this paper.

One possible candidate for a stable channel property is long-term average signal
strength, characterised by the average received power. This property can
have a significant impact on the \emph{inter-node} interference in massive MIMO,
particularly when simple maximum ratio combining (MRC) is used for precoding
\cite{marzetta2016fundamentals}. Furthermore, the average received power can be
effectively used for node grouping and scheduling \cite{fitzgerald2019massive}.
Nevertheless, inter-node interference can also be strongly affected by the
directional properties of the users' channels. Indeed measurement campaigns have
shown that propagation energy typically features dominant directions, and even
in reflective environments, real channel responses are far from ``ideal''
independent and identically distributed (i.i.d.) Rayleigh fading channels commonly
assumed in theory \cite{gao2015massivea, gao2015massiveb, Gunnarsson2018}. This
observation suggests that node partitioning that avoids simultaneously
scheduling nodes with propagation energy coming from similar directions can
actually prevent inter-node interference to a large extent\footnote{In terms of
    the ergodic sum spectral efficiency, naturally there is a price to pay to
    reduce inter-node interference through user grouping. However, given the low
    data rates required by typical IoT applications, energy efficiency is a more
    important performance metric for this use case than spectral efficiency.}.
    It is these solutions we investigate in this work.

\begin{figure}
    \begin{center}
	\includegraphics[width=0.7\columnwidth]{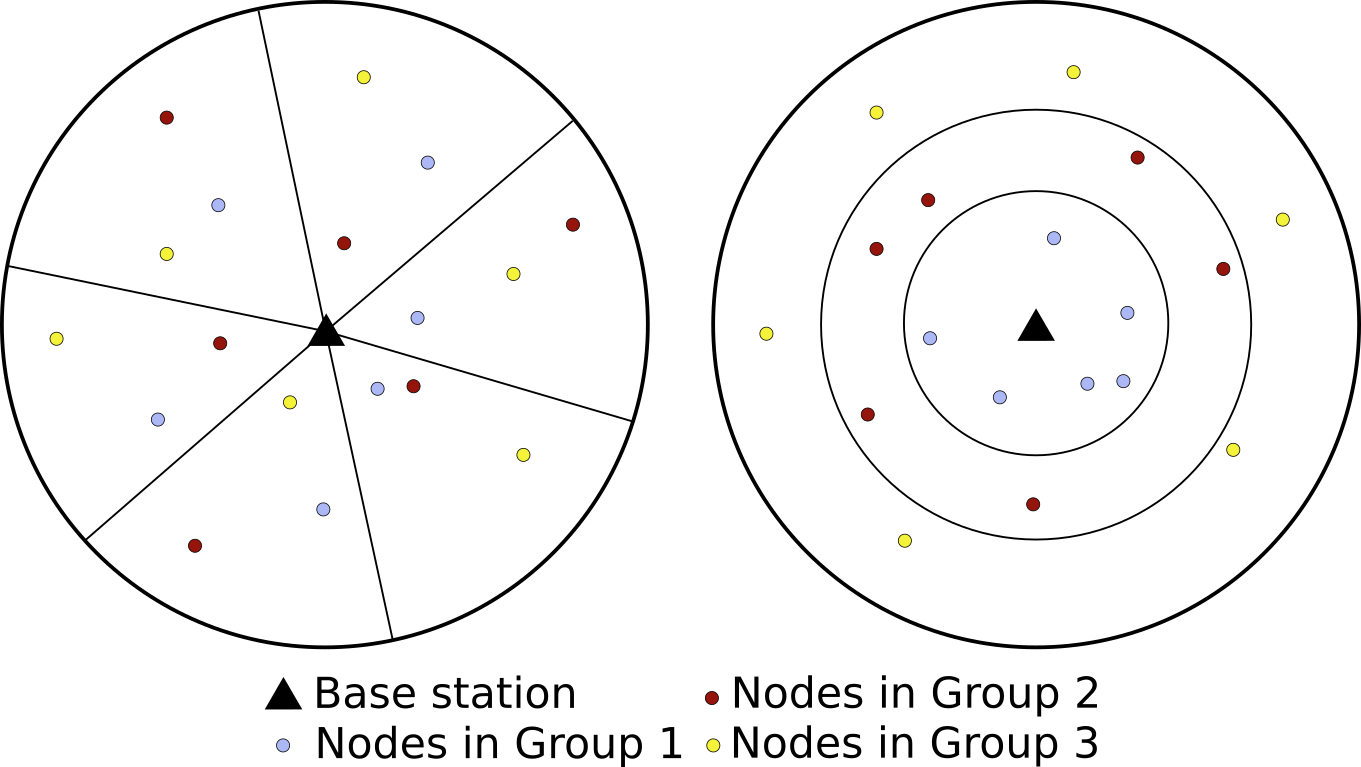}
	\caption{Two node partitioning approaches: ``pizza'' partitioning
	    (left), in which nodes are grouped such that the nodes in each group
	    have large differences in their directions from the base station;
	    and ``onion'' partitioning, in which nodes with similar signal
	    strength are grouped together. With line of sight and free space
	    propagation, shown here, the former is equivalent to dividing the
	    nodes into angular slices, while the latter is equivalent to
	    grouping the nodes into rings by distance from the base station.}
	\label{fig:pizza_and_onion}
    \end{center}
\end{figure}

We develop a node partitioning method for massive MIMO systems based on two
directional properties of the nodes' channels, namely dominant direction --- the
angle of arrival of the strongest channel component --- and angular spectrum
spread --- how widely the channel components are distributed over the possible
range of angles of arrival. These properties are stable over larger time scales
than the full CSI that changes according to the channel's coherence time, and are hence also robust against non-perfect CSI, easy
to collect at the base station with minimal signalling, and correlate with
inter-node interference. Specifically, we seek to partition the nodes so as to
maximize the minimum angular difference between nodes across multiple different
groups according to their angular spectrum spreads. 

This approach is analogous to partitioning the nodes within the coverage of a
centrally located base station into pizza slices, such that each slice only
contains one node from each group (Fig.~\ref{fig:pizza_and_onion}, left). This
can be compared to an onion ring partitioning approach, where nodes are grouped
such that nodes with similar signal strength are placed in the same group. With
line of sight and free space propagation, this would result in the groups
forming concentric shells, similar to the rings of an onion
(Fig.~\ref{fig:pizza_and_onion}, right). In more complex channel environments,
especially with shadowing or multipath components, the slices and rings will
not have such simple geometric shapes as shown in the figure and may even
consist of multiple disjoint regions. While we focus on the pizza approach in
this paper, with sufficiently many nodes, the two methods can be readily
combined, by simply composing them, for example dividing the nodes into onion
rings and then into pizza slices, or vice versa.

\section{Related Work}\label{sec:related_work}

Thus far, there has been limited work on node partitioning in massive MIMO in
the sense we consider in this paper. There is a substantial body of work on node
selection, starting with multiuser MIMO (MU-MIMO) systems
\cite{dimic2005downlink, wang2008user, huang2013user, shen2015sieve} even before
the advent of massive MIMO. However, the goal of these studies is to choose some
subset of the available nodes to be served at a given point in time, typically
to maximise the sum rate of the system. Although this is often referred to as
node scheduling, it is not true scheduling in that nodes that are not served
are not scheduled to be served in future: the nodes are rather partitioned into
those that are served, and those that are not, with the assumption that as
channel conditions change, different nodes will be served over time. We are
however interested in node partitioning for the purpose of then creating a
schedule for all node groups to be served over time. As such, multiple groups
are created, and all nodes must be included in at least one group for the
partition to be valid. To avoid confusion, we will therefore reserve the term
\emph{scheduling} to mean arranging service of nodes or node groups over time
such that all nodes or groups are eventually served, and use the term
\emph{selection} to mean selecting some subset of nodes for service at a given
point in time.

Furthermore, our objective is not to maximise the sum rate of the system. In many
IoT use cases, high throughput is not needed. Rather, the aim
is often to serve all node demands with minimal energy usage. Here, the rate
provided to an individual node does have some relation to this goal, since a
high achievable data rate can be used rather to reduce bit errors and thus
increase reliability, instead of maximizing throughput. Fewer bit errors mean
fewer retransmissions, and thus energy saved. Beyond a certain point, however,
nodes do not benefit from a higher SINR as they only have a small amount of
data, for example sensor or control data, to transmit or receive. As such, we
want to find a node grouping that maximizes the minimum SINR of any node, rather
than maximizing the sum rate.

Despite these differences, work on node selection is relevant to our work, since
a full node grouping can be regarded as an extension to the binary partitioning
of nodes into those that are served and those that are not, and can use
similar approaches. We will first discuss the extensive body of work on node
selection, and then consider the more limited existing work on scheduling and
random access in massive MIMO systems.

\subsection{Node Selection for Massive MIMO}

When selecting nodes for transmission, many approaches rely on full
CSI, such as \cite{tian2017overlapping, zhou2015joint,
benmimoune2015joint, guozhen2014joint}. This is however not suitable for our use
case since this would require all nodes to first wake up from sleep, and
then transmit a pilot signal in order for the base station to gather the CSI.
For low-power IoT devices, such an approach would consume too much energy to be
viable, and does not allow the schedule to be set in advance to provide a
defined sleep opportunity and wake up time.

There is however a body of work that performs node grouping based on reduced
channel information, which is usually more stable over time. This is often done
in the context of two-stage beamforming, where pre-beamforming groups are first
formed using coarse channel information, and then full precoding takes place
after collecting full CSI for a subset of nodes. Some examples of reduced
information that can be used for this purpose are channel quality indicators
\cite{lee2017new}, directional information collected from downlink probing
\cite{hajri2016scheduling}, and the covariance matrices of node channels
\cite{lee2017new,xu2014user,nam2014joint}, which can be determined over time by
recording statistical channel information. Of particular interest to us are
spatial clustering methods \cite{castaneda2016overview}, sometimes also used in
combination with the above techniques.

Joint spatial division and
multiplexing (JSDM) \cite{adhikary2013joint,nam2014joint} is one approach that has similarities to ours, although it is not aimed at node partitioning for scheduling. In this method,
artificial covariance matrices are constructed based on a set of angles of
arrival and angular spreads. Nodes are then clustered into sectors based on the
similarity of their channels, for example by computing the chordal distances of
the eigenspaces of nodes' channel covariance matrices to those of the
constructed ones. One key result, proved in \cite{adhikary2013joint}, is that
for a uniformly spaced linear array, JSDM approaches optimality, in terms of
maximizing the sum rate, as the number of antennas grows. This is because the
node channels, when grouped according to their angles of arrival and angular
spreads, form near-orthogonal subspaces. While the objective of JSDM is
different to that of our work, and thus results from JSDM and our method are not
comparable, this nonetheless demonstrates the value of node grouping based on
directional characteristics, which we also use in our approach.

A key difference between the above work and ours is that we do not wish to
select a subset of nodes for service to maximise the sum rate, but rather serve
all nodes over time. An optimal partitioning of nodes where only a subset will
be served may be quite different to one where all nodes must be served. In our
scenario, nodes with a low signal-to-noise ratio (SNR) or that interfere
significantly with other nodes cannot be simply dropped to improve overall
performance, but must be accommodated in some group.

A few existing works on node grouping consider other objectives. \cite{zhou2015joint}
groups nodes into multicast groups according to their interests, with the
objective of maximizing minimum throughput. However, here, full CSI is used,
which is not feasible for low-power IoT.  \cite{tian2017overlapping} focuses on
IoT, primarily considering the aspect of serving a large number of nodes, rather
than low energy usage, since again full CSI is used and the objective is maximal
sum rate. One interesting aspect of this work is that node groups may overlap,
such that each node can be served by more than one group, helping to ensure full
coverage. In our work, we use disjoint node groups, but this could be extended
to overlapping groups in the future, for example for nodes that have a higher
traffic demand or poor SINR.  \cite{lee2017new} performs node selection as in
the other work above, however this is done in combination with fair scheduling,
to avoid starvation of nodes that would otherwise be repeatedly not selected.
This still does not provide a predictable schedule that would allow a node to
sleep until its appointed service time, since nodes are still selected
dynamically in each time slot.

\subsection{Node Scheduling and Random Access for Massive MIMO}

There is also some, albeit limited, existing work on node scheduling for massive
MIMO. Because massive MIMO systems serve multiple nodes simultaneously, such
scheduling inherently requires creating node groups, although the groups do not
necessarily need to be disjoint. In \cite{fitzgerald2019massive}, joint node
scheduling and transmission power control is performed for nodes with
heterogeneous traffic demands. However, here, information about the traffic
queues at each node is required, which increases the signalling needed and thus
the energy cost. \cite{huh2011multi} examines fair scheduling in massive MIMO
systems, but provides an analysis of the achievable fair rate, rather than a
solution for actually performing scheduling.

Some work has also investigated random access for IoT devices in massive MIMO
systems \cite{liu2018sparse, de2017random,
bjornson2016random,sorensen2014massive, de2016random}. Random access is a
promising solution for IoT scenarios with sporadic and unpredictable traffic.
However, in cases where there is a high level of contention, random access can
lead to collisions, which induce retransmissions and cost energy while further
increasing the contention for the channel. Scheduled access also provides more
predictable performance, which can be important for some applications, and is
more suitable for predictable and/or periodic traffic.

Nevertheless, random access protocols could be combined with our node grouping
method, by allowing entire node groups to wake up during a limited period, in
which they would be served through random access. Using an intelligent node grouping
such as we propose then also reduces the collision probability within each group and 
increases the chances to distinguish colliding transmissions through techniques such
as successive interference cancellation. Alternatively, our node
grouping could be used as the precursor to any scheduling algorithm. In both
cases, node grouping simplifies network control, since devices can be
orchestrated at group level instead of addressing each individual device, and
provides for predictable and extended sleep opportunities.

\section{Scenario and Problem Definition}\label{sec:scenario}

We consider a single-cell network with the base station equipped with a large, central
antenna array, and a number of single-antenna IoT nodes. We here
consider static node devices, and they could be placed in different environments
(urban, indoors, rural, etc.), with the cell size varying according to the environment. Although the devices themselves do not move,
there may nonetheless be some movement in the environment, for example people
walking around.

We here give a brief overview of massive MIMO transmission underlying to our
work. For more details, the interested reader is referred to
\cite{marzetta2016fundamentals}. In a massive MIMO system, transmission is
established according to the coherence blocks of the channel. A coherence block
is a time by frequency region in which the channel is constant, to within some
small margin of error. Within each coherence block, nodes transmit pilot
signals in order for the base station to determine the CSI and thus allow data
transmission and reception. We consider a time-division duplex massive MIMO
system, relying on channel reciprocity within each coherence block, and hence
not requiring downlink pilots.

In each coherence block, a number of nodes are served simultaneously, for either
uplink or downlink transmission, or both. The number of nodes that may be served
in each coherence block is limited by either the number of available pilot
signals --- since each node must be allocated a pilot in order to transmit or
receive in the block --- or by the inter-node interference, or both. The
interference between nodes depends on both the combination of their channels and
the precoding scheme used. In this work, we focus on MRC as precoding as it can realize power efficient spatial multiplexing of devices~\cite{marzetta2016fundamentals}, based on the CSI that is acquired in the TDD-based massive MIMO operation.

In scenarios where the number of nodes exceeds the number that can be served
simultaneously in one coherence block, node grouping becomes necessary, such
that each group can be accommodated in a single coherence block. After
grouping, node groups can then be scheduled by assigning coherence blocks to
them such that the traffic demands of all nodes in the group are satisfied. In
this paper, we do not consider scheduling, but rather focus on the node
partitioning problem.

In this work, we perform node grouping based on two directional characteristics of the node
channels. The first is the dominant propagation direction, that is, the direction from which
the dominant component of a given node channel arrives at the base station. The
second characteristic we use is the angular spectrum spread: how much the power of the
node channel, as received by the base station, is spread over the angular
domain. A low angular spectrum spread thus implies a channel where the power is
largely focused in the dominant direction, whereas a large angular spectrum spread
indicates a channel with either multiple strong components, or a main component
that is spread over a wide angle, or both. Examples of channels with low and
high angular spectrum spreads are shown in Fig.~\ref{fig:low_and_high_spectral_spread}.

\begin{figure}
    \centering
    \resizebox{0.8\columnwidth}{!}{%
    \input{figures/min-max-spread-example.tex}}
    \caption{Two example channels with a high and low angular spectrum spread.
    Here, signal power is normalized such that the total power of each channel
across the entire angular spectrum is $1$.}%
    \label{fig:low_and_high_spectral_spread}
\end{figure}

For static nodes, these two characteristics of the channel will remain stable
over a long period of time, as demonstrated in~\cite{9539876,9443312}. If there is a line of sight component present, this
will typically constitute the dominant component, and its direction will not
change unless shadowing occurs to block it. Even in cases without a line of
sight component, there may be one or more strong multipath components, and
their directions will similarly be stable in the absence of variable shadowing. The
angular spectrum spread is influenced by the number and variety of multipath components: if there are many,
the channel is effectively ``spread out'' over a wider range of angles, leading
to a high angular spectrum spread. This means that even if the angles of arrival of
individual multipath channel components change, the overall angular spectrum spread will
again be relatively stable.

Not only are these directional characteristics stable over time, but they can
also be measured at the base station antenna array without requiring too much
complex processing~\cite{biswas2017new, 8486747, 9322427}. Moreover, the directional
properties of the nodes' channels are critical for the performance that can be
obtained, in terms of data rate and/or bit error rate. With MRC specifically,
the energy of a transmission (on the downlink) is focused at the target node,
however since this focusing is imperfect, especially in the presence of
imperfect CSI, interference is caused to nearby nodes. This means that
grouping nodes in such a way as to spread out the nodes in angle leads to
improved SINRs for the nodes.  For these reasons, node partitioning based on
directional channel characteristics has the potential to provide a simple
partitioning method with minimal information collected from node devices, thus
allowing them to sleep longer, transmit less, and save energy.

\section{Optimization}\label{sec:system_model}

We will now present the node grouping optimization problem. We begin by
detailing the system model and notation, and will then describe our
mixed-integer programming optimization formulation. The notation is summarised
in Table~\ref{tab:notation}.

\begin{table}
    \centering
    \renewcommand{\arraystretch}{1.2}
    \caption{Notation}
    \begin{tabularx}{\columnwidth}{c X}
    	\toprule
	$\mathcal K$ & set of nodes\\
	$\theta(k)$ & dominant direction of node $k \in \mathcal K$ \\
	$P$ & number of pilots in each coherence block \\
	$\mathcal G$ & set of scheduling groups \\
	$u_{kg}$ & whether or not node $k \in \mathcal K$ is placed in group $g
	\in \mathcal G$\\
	$\mathcal P$ & the set $\{1, ..., P\}$, $P \in \mathbb N$\\	
	$Y_{kg}^p$ & whether or not node $k \in \mathcal K$ is the $p$th node
	in group $g\in \mathcal G$, $p\in \mathcal P$\\
	$t_g^p$ & angle of the $p$th node in group $g \in \mathcal G$, $p
	\in\mathcal P$\\
	$T_g$ & angle of the last node in group $g \in \mathcal G$ \\
	$d_g^p$ & $p$-th angular difference of nodes in group $g\in \mathcal G$,
	$p \in \mathcal P$\\
	$B$ & min-max angular difference across groups \\
	$\sigma(k)$ & angular spectrum spread of node $k\in\mathcal K$\\
	$s_g^p$ & angular shift of the $p$th node in group $g \in \mathcal G$,
	$p \in \mathcal P$\\
	$S_g$ & angular shift of the last node in group $g \in \mathcal G$\\
	$m_g^p$ & whether or not the $p$th node in group $g \in \mathcal G$ is
	the last node in the group, $p \in \mathcal P$ \\
	\bottomrule
    \end{tabularx}
    \label{tab:notation}
\end{table}

\subsection{System Model}

We have a set of nodes (IoT devices) $\mathcal K$ served by a massive MIMO
base station. Each node $k \in \mathcal K$ has a dominant direction of its signal as
defined in the previous section, denoted by $\theta(k)$, and measured in radians
clockwise from a designated reference direction (e.g., due north). In the
following, if not otherwise specified, all other angles are also defined as
clockwise from this reference direction. The dominant direction does not
necessarily represent the direction to the physical location of the node; it
could also be a signal component reflected one or more times from an object in
the environment. For our purposes, however, the physical
location of the node is less important than the arrival direction of its
transmitted signals.

Scheduling of nodes for both uplink and downlink transmission occurs in
coherence blocks, such that a given group of nodes should be scheduled for an
integer number of blocks. We will here focus on partitioning the set of nodes $\mathcal K$ into groups that can be
scheduled together. In each coherence block, there are $P$ available pilot
signals, and this thus represents the maximum number of nodes that can be placed
in each group.

We define a set of scheduling groups $\mathcal G$, where each group may contain
at most $P$ nodes. The groups do not necessarily need to contain the same number
of nodes, and some groups may even be empty. The partitioning of nodes into
groups is represented by binary decision variables $u_{kg}$, $k \in \mathcal K$,
$g\in \mathcal G$, equal to $1$ if and only if node $k$ is placed in group $g$.
Within each group, the nodes are ordered by their dominant directions. We
therefore define binary variables $Y_{kg}^p$, $k\in \mathcal K$, $g\in \mathcal
G$, $p \in \mathcal P = \{1, \dots, P\}$, where $Y_{kg}^p = 1$ indicates that
$k$ is the $p$th node in $g$. If a given group $g$ contains fewer than $P$
nodes, some $Y_{kg}^p$ will be zero for all $k \in \mathcal K$. The angle of the
$p$th node in group $g$ is given by the continuous variable $t_g^p$, with $t_g^p
= 0$ if group $g$ contains fewer than $p$ nodes.

We next introduce continuous variables $d_g^p$, $g\in \mathcal G$, $p \in
\mathcal P$ to represent the differences in angle between adjacent nodes in each
group. For $p$ less than the number of nodes in group $g$, we thus have $d_g^p =
t_g^{p+1} - t_g^p$ for $p < P$. Variable $d_g^P$ is used to represent the
angular difference between the last and the first nodes node in group $g$ ---
thus completing the circle --- regardless of the total number of nodes in the
group, that is, even if group $g$ has fewer than $P$ nodes.  Finally, we define
the continuous variable $B$, which represents the maximum of the minimum angular differences
in all the groups, and will constitute our objective function. 

In addition to the dominant direction, we use the angular spectrum spread to
inform node grouping (see Section~\ref{sec:scenario}). In massive MIMO systems,
the signal from a node with a large angular spectrum spread is more easily
distinguished from other signals, even from nodes that have similar dominant
directions, since there are more multipath components to use to differentiate
the channels. This means that a node with a large angular spectrum spread causes
less interference to nearby (in angle) nodes, and so can be more readily placed
in a group with them. 

For the purposes of partitioning the nodes into scheduling groups, we model
the angular spectrum spread by considering an angular tolerance allowing nodes
to symbolically ``move'' around a circle from their initial positions (given by
their dominant directions), in such a way as to increase their angular
differences to the neighbouring nodes in the group. We first define the
normalised angular spectrum spread of a node $k \in \mathcal K$, $\sigma(k)$,
where $0\leq \sigma(k) \leq 1$. Here, $\sigma(k) = 0$ means the node's signal is
entirely composed of a component coming from its dominant direction, with no
angular spectrum spread. In such a case, the node will not be permitted to move
from its initial direction $\theta(k)$. On the other hand, $\sigma(k) = 1$ means
that the dominant direction cannot be distinguished at all: the node's signal
comes evenly from all directions, i.e. we have an i.i.d. Rayleigh fading
channel. Such a node may symbolically move freely to any angle around the
circle, and this shift is represented by the decision variable $s_g^p$, for the
$p$th node in group $g \in \mathcal G$. Of course, in reality the nodes do not
move; shifting the nodes on the circle corresponds rather to reducing the
influence of the angular proximity of two nodes if their angular spectrum
spreads allow their channels to be distinguished by the base station anyway,
equivalent to if they had a larger angular difference in the first place.

Lastly, we will use decision variables $m_g^p$ to indicate whether or not the
$p$th node in group $g \in \mathcal G$ is the last node in this group. This will
allow us to compute variables $T_g$ and $S_g$, the angle and angular shift of
the last node in each group $g \in \mathcal G$, respectively. These will be used
to correctly compute the angular differences to neighbouring nodes, since the
last node in each group is a special case, as one of its neighbours is the first
node in the group as the nodes are placed on a circle.

\subsection{Formulation}

We can now formulate the optimization problem as follows.

\begin{subequations} \label{form:pizza}
    \begin{align}
	\max \quad & B & \label{eq:obj} \\
	& \sum_{g \in \mathcal G} u_{kg} = 1, & k \in \mathcal K
	\label{eq:one_group} \\
	& \sum_{k \in \mathcal K} u_{kg} \leq P, & g \in \mathcal G
	\label{eq:num_pilots} \\
	& \sum_{p \in \mathcal P} Y_{kg}^p = u_{kg}, & k \in \mathcal K, \, g \in
	\mathcal G \label{eq:node_order} \\
	& \sum_{k \in \mathcal K} Y_{kg}^p \leq 1, & p \in \mathcal P, \, g \in
	\mathcal G \label{eq:one_node_per_position} \\
	& \sum_{k \in \mathcal K} Y_{kg}^p \leq \sum_{k \in \mathcal K}
	Y_{kg}^{p+1}, & p \in \mathcal P \setminus \{P\}, \, g \in \mathcal G 
	\label{eq:no_gaps} \\
	& t_g^p = \sum_{k \in \mathcal K} \theta(k) Y_{kg}^p, & p \in \mathcal
	P, \, g \in \mathcal G \label{eq:select_angle} \\
	& t_g^p \leq  t_g^q + 2\pi\left(1 - \sum_{k \in \mathcal K}
    Y_{kg}^p\right), & p, \, q \in \mathcal P, \, p < q, \, g \in \mathcal G
	\label{eq:angle_order} \\
	& s_g^p \leq \pi \sum_{k \in \mathcal K} \sigma(k) Y_{kg}^p, & p \in
	\mathcal P, \, g \in \mathcal G \label{eq:angular_shift} \\
	& T_g \geq t_g^p, & p \in \mathcal P, g \in \mathcal G
	\label{eq:last_angle} \\
	& T_g \leq \sum_{p \in \mathcal P} t_g^p m_g^p, & g \in \mathcal G
	\label{eq:T_g} \\
	& S_g = \sum_{p \in \mathcal P} s_g^p m_g^p, & g \in \mathcal G
	\label{eq:S_g} \\
	& \sum_{p \in \mathcal P} m_g^p = 1, & g \in \mathcal G
	\label{eq:select_one} \\
	& d_g^p \leq t_g^{p+1} - t_g^p + s_g^{p+1} + s_g^p, & p \in \mathcal P \setminus \{P\}, \, g \in \mathcal G
	\label{eq:angular_diff} \\ 
    & d_g^P \leq t_g^1 + 2\pi - T_g + s_g^1 + S_g, & p \in \mathcal P \setminus \{P\}, \, g \in \mathcal G
	\label{eq:angular_diff_last} \\
	& d_g^p \leq \pi, & p \in \mathcal P, \, g \in \mathcal G
	\label{eq:angular_diff_pi} \\
	& \mathrlap{B \leq d_g^p + \pi\left(1 - \sum_{k \in \mathcal K}
	Y_{kg}^p\right),} & g \in \mathcal G \label{eq:min_ang_diff} \\
	& B \leq d_g^P, & g \in \mathcal G \label{eq:min_ang_diff_last} \\
	& u_{kg}, \, Y_{kg}^p, \, m_g^p \in \mathbb B; \;
	t_g^p, \, s_g^p, \, T_g, \, S_g \in \mathbb R; \; 
	B, \, d_g^p \in \mathbb R^+, & p \in \mathcal P, \, g \in \mathcal G.
    \end{align}
\end{subequations}

The objective function, \eqref{eq:obj}, maximizes the minimum angular
difference, adjusted for angular spectrum spread, between any two nodes in any
one group, based on their symbolic placement around a circle. We assume that
performance, in terms of achievable SINR, will in general improve for any given
node the greater the angle between it and its closest neighbours that are
scheduled in the same group. However, this relationship is not straightforward
for channels with multipath components, since nodes whose dominant directions
lie close to one another may nonetheless have little correlation between their
channels if they have significant power in multipath components with different
angles of arrival. The minimum angular difference that is maximized therefore
takes the angular spectrum spread of the nodes' channels into account, via the
constraints that will be explained below. 

The first constraint, \eqref{eq:one_group}, ensures that each node is placed in
exactly one group, and constraint \eqref{eq:num_pilots} then ensures that the
number of nodes placed in each group does not exceed the number of available
pilots, $P$. The decision variables $Y_{kg}^p$ are used to keep track of which
group each node is placed in along with in which order. Hence $Y_{kg}^p$ should
be set to $1$ if node $k \in \mathcal K$ is placed in group $g \in \mathcal G$
at position $p \leq P$, and $0$ otherwise. This is enforced using constraints
\eqref{eq:node_order} and \eqref{eq:one_node_per_position}. Constraint
\eqref{eq:node_order} requires that exactly one position in group $g \in
\mathcal G$ be assigned to node $k \in \mathcal K$, if the node is placed in
that group ($u_{kg} = 1$), and that no positions in the group are assigned to it
if it was not placed in that group ($u_{kg} = 0$). Constraint
\eqref{eq:one_node_per_position} ensures that at most one node is assigned to
each position in each group, i.e. a strict ordering is imposed in which two or
more nodes cannot occupy the same position. Finally, constraint
\eqref{eq:no_gaps} prevents gaps in the node ordering for each group, that is,
no position in the group may be filled until all the preceding positions are
filled. 

The next three constraints determine the angles and angular tolerances of each
node in each group. Constraint \eqref{eq:select_angle} sets the angle of each
node position in each group to be the dominant direction of the node assigned
that position, through the decision variable $t_g^p$. Constraint
\eqref{eq:angle_order} ensures that nodes are assigned positions in groups in
order of their dominant directions, that is, a node with a higher angle for its
dominant direction may not be placed earlier in a group than a node with a lower
angle. The second term on the right-hand side is used to cancel this constraint
when a position in a group is empty, that is, not assigned to any node. Such
empty positions must occur at the end of the group, thanks to constraint
\eqref{eq:no_gaps}. Constraint \eqref{eq:angular_shift} determines the angular
tolerance for each position in each group, up to a maximum of the angular
spectrum spread for the node assigned that position. This angular tolerance
gives the range within which the node's angular position can be adjusted.

Because the nodes are symbolically placed on a circle, in which an angle of $2\pi$ is
equivalent to an angle of $0$, the last node in each group must be treated
differently. For the purpose of computing the objective, angular differences
between each successive pair of nodes in each group are considered, and
additionally the angular difference from the last node back around the remainder
of the circle to the first node must be included. Constraints
\eqref{eq:last_angle}--\eqref{eq:select_one} are used to find the angle and
angular shift for the last node in each group. The binary decision variable
$m_g^p$ is set to $1$ for the last position in each group, and $0$ otherwise. It
is not possible to determine in advance which is the last position in any given
group, since the groups may have different numbers of nodes assigned to them.

Constraint \eqref{eq:select_one} forces only one position in each group to be
selected as the last. To make sure the one selected is indeed the last, and
obtain the angle of the node in the last position, constraints
\eqref{eq:last_angle} and \eqref{eq:T_g} are used. Constraint
\eqref{eq:last_angle} requires that the angle of the last position in each group
$g \in \mathcal G$, $T_g$, be at least as large as all angles in the group
$t_g^p$. Meanwhile, constraint \eqref{eq:T_g} requires that $T_g$ be no more
than the angle of the node selected by the variable $m_g^p$. In this way,
$m_g^p$ can only feasibly be $1$ for the last position in the group. Similarly,
variable $S_g$ is set to the angular shift of the last node in the group via
constraint \eqref{eq:S_g}.

Next, the angular differences between the nodes in each group need to be
computed, in constraints \eqref{eq:angular_diff}--\eqref{eq:angular_diff_pi}.
Constraint \eqref{eq:angular_diff} sets variable $d_g^p$ to at most the angular
difference between the nodes at positions $p$ and $p+1$ in group $g$, taking
into account their angular tolerances. Since the objective is a maximization
function, equality will be achieved here for any angular differences that can
affect the objective. Constraint \eqref{eq:angular_diff_last} similarly takes
the angular difference between the last and first positions in each group, thus
closing the circle. Note that the variable $d_g^P$ is used for this last angular
difference, regardless of whether or not the last occupied position in the group
is $P$ or a lower position, due to the group having fewer than $P$ nodes
assigned to it.  Constraint \eqref{eq:angular_diff_pi} limits all angular
differences to be at most $\pi$, since any difference greater than this would
imply a smaller angular difference if measured in the opposite direction around
the circle.

We are now ready to compute the objective function using constraints
\eqref{eq:min_ang_diff} and \eqref{eq:min_ang_diff_last}. These constraints set the
objective function value, represented by variable $B$, to the minimum angular
difference of any pair of nodes in any group. Here, again, the last position in
each group must be treated separately (constraint \eqref{eq:min_ang_diff_last}). In
constraint \eqref{eq:min_ang_diff}, the second term on the right-hand side cancels
the constraint for any positions to which no node has been assigned. Finally,
the last three constraints in the formulation provide the domains for each of
the decision variables.

\begin{figure}
    \begin{center}
	\includegraphics[width=0.4\columnwidth]{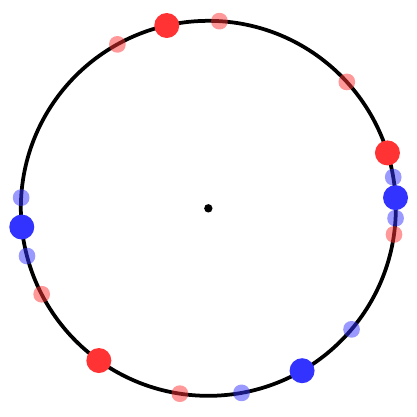}
	\caption{Example problem instance and solution for the optimization
	    problem. Six nodes' dominant directions (larger circles) and maximal
	    allowed angular shifts (smaller circles) are shown placed on a
	    circle around the base station (central black circle). The nodes are
	    partitioned into two groups, red and blue, such that the minimum
	    angular difference between any two nodes is maximized, after allowing
	    nodes to shift to any position within the range given by their maximal
	    allowed angular shifts.}
	\label{fig:opt_example}
    \end{center}
\end{figure}

An example problem instance and solution are shown in
Fig.~\ref{fig:opt_example}. In this instance, six nodes are partitioned into
two groups. The nodes' angular shift ranges are indicated in the figure by the
smaller, semi-transparent circles. For this problem, most nodes' ranges do not
overlap and so do not affect the solution: the nodes are placed into the two
groups in an alternating fashion. However, two of the nodes' ranges overlap, and
in this case the nodes are grouped ``out of order'', as they are able to
shift within their ranges in order to achieve bigger angular differences to the
other nodes.

\subsubsection{Auxiliary Variables}

To resolve the bilinearities in constraints \eqref{eq:T_g} and
\eqref{eq:S_g}, we need to introduce 
auxiliary variables $v_g^p = t_g^p m_g^p$ and $z_g^p = s_g^p m_g^p$, $p \in
\mathcal P$, $g \in \mathcal G$, and add the following constraints to
formulation~\eqref{form:pizza}.
\begin{subequations}
    \begin{align}
	& v_g^p \leq t_g^p; \; 
	v_g^p \leq 2\pi m_g^p; \; 
	v_g^p \geq t_g^p + 2\pi\left(m_g^p - 1\right), & p \in
	\mathcal P, \, g \in \mathcal G \\
	& z_g^p \leq s_g^p; \; 
	z_g^p \leq \pi m_g^p; \; 
	z_g^p \geq s_g^p + \pi\left(m_g^p - 1\right), & p \in \mathcal P, \, g \in \mathcal G.
    \end{align}
\end{subequations}
Constraint~\eqref{eq:T_g} is then replaced with $T_g \leq \sum_{p \in \mathcal
P} v_g^p, \, g \in \mathcal G$,
and constraint~\eqref{eq:S_g} with $S_g = \sum_{p \in \mathcal P} z_g^p, \, g \in
\mathcal G$.
This renders the formulation as a valid mixed-integer programming problem, but
is otherwise equivalent to formulation \eqref{form:pizza}.

\section{Approximation Algorithm}\label{sec:approximation}

Solving formulation \eqref{form:pizza} to optimality may require more time than
is available to partition the nodes and schedule their transmissions. As we will
see in Section~\ref{sec:evaluation_results}, the time to find an optimal
solution increases exponentially with the size of the problem, and for many IoT
scenarios the number of nodes can be very large. In some cases, it may
nonetheless be possible to use optimal solutions, since it is also often the
case that IoT devices transmit only infrequently, for example sensor data
gathered once every hour or day, which, depending on the number of nodes, may
allow enough time to solve the optimization problem.

In many cases, there will not be sufficient time and/or
computational resources available or justifiable to find an optimal solution
before it is needed, especially in view of the current trend of moving more
intelligence deeper in the network. To address such scenarios, we created a
suboptimal algorithm based on the same principles as the optimization
formulation. This algorithm is efficient and can be used even for scenarios with
many nodes that need to be partitioned and scheduled frequently. The algorithm
is listed in Algorithm \ref{alg:approx}. Since the domain of our optimisation
(the circle) is finite, there is an inherent additive bound on the solution as
compared with the optimal solution: the deviation of the solution provided by 
Algorithm~\ref{alg:approx} from the optimal solution cannot be greater than $\pi$, and hence
we refer to it as an approximation algorithm in the following.

\IncMargin{1em}
\begin{algorithm}
    \small
    \KwIn{$\mathcal K$: set of nodes, each with dominant direction $\theta(k)$
    and angular spectrum spread $\sigma(k)$\\
    \quad\quad\quad$\mathcal G$: set of groups numbered $1\dots G$, $G=|\mathcal G|$}
    \KwOut{$group$: list of groups each node is assigned to, of length
    $|\mathcal K|$}
    \ForEach{$k \in \mathcal K$}
    {
	$min\_angle[$k$] \leftarrow \theta(k) - \pi\sigma(k)$\\
	$max\_angle[$k$] \leftarrow \theta(k) + \pi\sigma(k)$\\
    }
    $sorted\_nodes \leftarrow \mathcal K$ sorted by $min\_angle$\\
    $first\_node \leftarrow$ pop($sorted\_nodes$)\\
    $group[first\_node] \leftarrow 1$\\
    $next\_group \leftarrow 2$\\
    $sorted\_nodes \leftarrow \mathcal K$ sorted by $max\_angle$\\
    \ForEach{$k \in \mathcal K$}
    {
	$group[k] \leftarrow next\_group$\\
	$next\_group \leftarrow next\_group + 1$\\
	\If{$next\_group > G$}
	{
	    $next\_group \leftarrow 1$\\
	}
    }

    \caption{Approximation algorithm for node partitioning based on dominant
    direction and angular spectrum spread}
    \label{alg:approx}
\end{algorithm}

The basic idea of the approximation algorithm is to sort the nodes by angle,
that is by their dominant directions but also taking into account their angular
spectrum spreads, and then assign a group to each node sequentially in a round
robin fashion. In this way, nodes in the same group will be as far away as
possible from each other in terms of their position in the sequence, although
not necessarily in terms of their angles since the solutions produced by this
algorithm are suboptimal in general.

To take the angular spectrum spreads into account, an angular shift is computed for each
node $k \in \mathcal K$ as $\pi \sigma(k)$, where $\sigma(k)$ is the normalised
angular spectrum spread of $k$. This is similar to constraint \eqref{eq:angular_shift}
in the optimization. The angular shift is first subtracted from each node's
dominant direction (line $2$), and the node with the minimal resulting angular
position is taken as the first node and assigned to the first group (lines
$5$--$6$).  Then, for the remaining nodes, the angular shift is added to each
node's dominant direction (line $3$) and the nodes are again sorted by the
resulting angle (line $9$). Nodes are taken from the head of the resulting list
one at a time in order and assigned a group in a round robin fashion (lines
$10$--$15$), that is, cycling through the groups in order and assigning one node
to each before repeating the process.

The most complex operation in the algorithm is sorting the nodes, which is done
twice sequentially. Efficient sorting algorithms such as quick sort and merge
sort can perform this operation in $O(KlogK)$ time, where $K = |\mathcal K|$ is
the number of nodes to be sorted. The two sequential for loops each have a complexity
of $O(K)$, which is lower than that of the sorting operation so it does not
increase the overall complexity class. All other operations are basic operations that
are executed in $O(1)$ time. Thus the total computational complexity of the
approximation algorithm is $O(KlogK)$. 

It should also be noted that in cases where the angular spectrum spreads are low, such
that the resulting angular shifts are lower than the angular differences between
nodes' dominant directions for most nodes, the list of nodes will already be
nearly sorted, assuming it is initially provided in order of nodes' dominant
directions. In this case, the complexity could be reduced even further by using
bubble sort, which has good performance of $O(K)$ for an almost-sorted list
where no node is out of place by more than one position. In such a case, the
complexity of our approximation algorithm also becomes $O(K)$. Since node
partitioning based on angles, as proposed in this paper, is intended for cases
where node channels have a strong dominant component and thus interference
correlates strongly with the directional characteristics of the channel, cases where
the angular spectrum spread is low enough to use bubble sort are quite likely to occur.
Checking whether a given problem instance fulfills the needed conditions can be
performed in $O(K)$ time, for example by performing the first pass of bubble
sort and, in case the list is not yet sorted, switching to quick sort or merge
sort. Thus, this added optimization for such cases can be included at no
additional cost in terms of computational complexity.

\section{Numerical evaluation}\label{sec:evaluation}

We conducted a numerical evaluation to test the performance of our optimization
and approximation algorithm, along with two reference algorithms, one of which uses 
full CSI, on randomly
generated instances of our problem scenario, consisting of a set of
single-antenna nodes and an accompanying channel matrix. We compared the
performance of the algorithms in terms of the achieved minimum, maximum, and
average SINR. We also compared the performance of our approximation with the
optimal solution in terms of the proportion of the optimal objective function
value obtained by the approximation. All source code for our implementations of
the optimization problem, approximation algorithm, channel model, and
experiments is available online \cite{pizza}.

We generated instances consisting of between $15$ and $36$ nodes, along with a
massive MIMO base station with $100$ antennas, and $12$ pilots available in each
coherence block. These base station parameters are based on the Lund University massive
MIMO (LuMaMi) testbed \cite{vieira2014flexible,malkowsky2017world}. The number of nodes and pilots were also chosen such that there are always fewer pilots than nodes, necessitating node grouping.

\begin{table}
    \begin{center}
	\caption{Parameters for the numerical evaluation.}%
	\label{tab:parameters}
	\begin{tabularx}{\columnwidth}{X l c}
	    \toprule
	    \textbf{Parameter} & \textbf{Symbol} &\textbf{Value} \\
	    \midrule
	    Number of nodes & $|\mathcal{K}|$ & $15\dots36$, step $3$ \\
	    Number of instances for each number of nodes & -& $20$ \\
	    Number of pilots & $P$ & $12$ \\
	    Number of groups & $|\mathcal{G}|$ & $3$ \\
	    Number of base station antennas & $M$ & $100$ \\
	    Type of antenna array &-& Uniform rectangular array \\
	    Precoding &-& Maximum ratio combining \\
	    Carrier frequency &-& \SI{2.47}{\giga\hertz}\\
	    Number of channel clusters &$N_{\mathrm{cl}}$ & $4$ \\
	    Number of channel components per cluster & $N_\mathrm{ray}$ & $10$ \\
	    Distribution of cluster central angles & -& Uniform $45$--$135$ degrees \\
	    Distribution of component angles within each cluster (vertical and
	    horizontal)  & -& Exponential, $\mu=7.5$ degrees \\
	    Component small scale fading distribution  & -& Complex normal, $\mu=0$,
	    $\sigma=1$ \\
	    Signal-to-noise ratio (for all nodes)  & -& $20$ dB \\
	    \bottomrule
	\end{tabularx}
    \end{center}
\end{table}

We partitioned the nodes into three groups, and so as the number of nodes
increased, the number of nodes per group increased accordingly. The number of
nodes per group is the most important parameter for determining the achievable
SINR, since with more nodes per group the density of nodes is greater and the
angular differences between them are lower. Meanwhile, the overall number of
nodes is the most important parameter in terms of the running time of the
partitioning algorithms.

For each number of nodes, we generated and tested $20$ problem instances. The
instances were generated using a channel model based on the Saleh-Valenzuela
model~\cite{saleh1987statistical} and its adaption to MIMO
channels~\cite{el2014spatially}. The channel model and problem instance
generation will be further detailed in Section~\ref{sec:channel_model}. As
output, the model produces the dominant direction and angular spectrum spread
for each node, along with the complete channel matrix giving the channel from
each node to each base station antenna.

Using the dominant directions and angular spectrum spreads, we then ran our optimization
algorithm, given in formulation~\eqref{form:pizza}, on each problem instance.
For this we used the AMPL modelling language~\cite{ampl2002} along with the
CPLEX optimization solver~\cite{cplex}, and the optimization was run on a
12-core server with $2.2$ GHZ Intel Xeon ES-2420 CPUs and $24$ GiB RAM. The
optimal solution was then compared with three other node partitioning
algorithms. An overview of the partitioning algorithms is given in
Table~\ref{tab:partitioning-algorithms}. The first comparison algorithm was our
approximation algorithm, given in Algorithm \ref{alg:approx}.

\begin{table}[h]
    \centering
    \caption{Partitioning Algorithms Tested.}%
    \label{tab:partitioning-algorithms}
    \begin{tabularx}{\columnwidth}{l X}
    \toprule
         Node Partitioning Method &  Description \\
         \midrule 
         Optimal Solution & Solving formulation~\eqref{form:pizza} to optimality.\\
         Approximation algorithm &  Algorithm~\ref{alg:approx}\\
         Clumped partitioning & Groups nodes with similar directional
         properties together. Worst-case scenario using directional properties.\\
         Power partitioning &  Groups nodes with similar received power at the
         base station together. Partitioning suitable for use with max-min fair
         power control. Requires full CSI. \\
    \bottomrule
    \end{tabularx}
\end{table}

The second comparison algorithm, which we will refer to as \emph{clumped} node
partitioning, is intended to represent a worst-case scenario in which nodes
whose channel have similar directional properties are grouped together. In this
algorithm, the nodes are first sorted according to their dominant directions.
Then the first $N = \frac{|\mathcal K|}{|\mathcal G|}$ nodes are placed in the
first group, the next $N$ nodes in the next group, and so on until all nodes
have been assigned a group. The computational complexity of this algorithm is $O(log(|\mathcal K|)$, since it first sorts the nodes then performs a sequential group assignment for each node. In our test instances, the number of nodes is always
divisible by the number of groups, so there will be no nodes left over with this
procedure, and all groups will contain the same number of nodes. Clumped
partitioning gives an indication of the detriment to performance that might be
caused by a poor node partition with respect to the directional channel
properties we consider.

The third comparison algorithm, called \emph{power} partitioning, uses the
entire channel matrix as input, rather than only the directional properties of the
node channels, and thus is used as a comparison against the case where full
channel information is obtained before partitioning. To the best of our knowledge, no 
existing methods from the literature take a similar approach to ours of avoiding the collection
of CSI and instead using other properties. Collection of full CSI is not feasible in
practice. This is firstly because it would require partitioning nodes after they
have transmitted pilots in the coherence block in which they are to transmit,
whereas in reality pilots can only be allocated to nodes after partitioning has
been performed. Secondly, in an IoT scenario, the signalling cost for collecting
channel information in order to perform partitioning is too high; our aim is to
provide a partitioning strategy with low energy usage on the end devices.

Power partitioning sorts the nodes according to the received power of their
channels at the base station, and then groups those nodes with similar power
levels. Similarly to clumped partitioning, this is done by assigning the $N =
\frac{|\mathcal K|}{|\mathcal G|}$ nodes with the highest received power to the
first group, the next $N$ highest power nodes to the next group, and so on. As for clumped partitioning, the computational complexity is $O(log(|\mathcal K|)$, as power partitioning works in a similar way. This does not include the complexity for collecting and processing the CSI in order to find the received power for each node, since this will depend on the specific implemention for downlink precoding and uplink combining. For
MRC precoding, the interference scales with the interfering nodes' received
power~\cite{marzetta2016fundamentals}. Power partitioning thus provides a
reasonably fair partition in which nodes experience interference only from other
nodes with similar received power. 

Power partitioning is thus similar to max-min fair power control \cite{bjornson2016massivea}, in
which a common SINR is achieved for all users by adjusting their transmission
power such that the received power from each user at the base station is equal.
In the absence of transmission power control, fair performance --- in the form
of equal SINR --- cannot be achieved exactly, however power partitioning comes as
close as possible using only partitioning. If partitioning were to then be
combined with transmission power control, power partitioning would then provide
the best conditions for carrying out max-min fair power control since the users
in each group will have as similar received power levels as possible before
applying the power control algorithm. For these reasons, power partitioning
provides a good comparison algorithm making use of full CSI for our use case, in
which the goal is to maximise the minimum SINR, rather than provide the maximum
rum rate as in other user selection algorithms (see
Section~\ref{sec:related_work}).

Finally, we computed the SINR achieved for each node for the partitions produced
by each of the algorithms. Here, the interfering nodes are
only those that were placed in the same group, as it is these nodes that will
transmit in the same coherence block. We then took the minimum SINR obtained by
any node in any group as our primary performance metric, however we also
examined the average and maximum SINR across all the nodes.

\subsection{Channel Model}\label{sec:channel_model}

To generate each problem instance, we used the channel model
from~\cite{el2014spatially}, which is in turn based on the Saleh-Valenzuela
model~\cite{saleh1987statistical}. The model defines the channel matrix $\textbf{H}$ as
\begin{equation}
    \textbf{H} = \sqrt{%
    \frac{M}%
    {N_{\text{cl}} N_{\text{ray}}}%
    }
    \sum^{N_{\text{cl}}}_{i=1} \sum^{N_{\text{ray}}}_{l=1}%
    \alpha_{i \ell} \Lambda_{\mathrm{r}}\left(\phi_{i \ell}^{\mathrm{r}}, \theta_{i \ell}^{\mathrm{r}}\right) \Lambda_{\mathrm{t}}\left(\phi_{i \ell}^{\mathrm{t}}, \theta_{i \ell}^{\mathrm{t}}\right)\textbf{a}_\mathrm{r}\left(\phi_{i \ell}^{\mathrm{r}}, \theta_{i \ell}^{\mathrm{r}}\right).
\end{equation}

The channel matrix $\textbf{H}$ models the effect of all the nodes' signals arriving at the $M$ receive antennas of the base station. At the nodes' side, single antennas are considered. The signals are modeled as multiple components grouped into $N_{\mathrm{cl}}$ different clusters. 
Clusters can for example be created by objects in the environment that the nodes'
signals are reflected from (possibly multiple times) before arriving at the
base station. Each cluster has a central angle-of-arrival in relation to the base station,
around which its $N_{\mathrm{ray}}$ constituent components propagate. 
The propagation effects (attenuation and phase rotations) are modeled in the complex gain $\alpha_{i \ell}$ for each contributing ray. The receive array response is denoted by $\textbf{a}_\mathrm{r}\left(\phi_{i \ell}^{\mathrm{r}}, \theta_{i \ell}^{\mathrm{r}}\right)$ and is a function of the antenna array structure, the azimuth, and elevation angles ($\phi_{i \ell}^{\mathrm{r}}, \theta_{i \ell}^{\mathrm{r}}$) of the arriving signal of the $\ell^{\text{th}}$ ray in the $i^{\text{th}}$ cluster. The expression of the array response for a uniform linear array (ULA) and a uniform rectangular array (URA) can be found in~\cite{balanis2015antenna}. The antenna element gain of the receive and transmit antennas are denoted by the functions $\Lambda_{\mathrm{r}}\left(\phi_{i \ell}^{\mathrm{r}}, \theta_{i \ell}^{\mathrm{r}}\right)$ and $\Lambda_{\mathrm{t}}\left(\phi_{i \ell}^{\mathrm{t}}, \theta_{i \ell}^{\mathrm{t}}\right)$, respectively. For isotropic antennas, the antenna element gain is one ($\Lambda=1$), as the gain is independent of the azimuth and elevation angle.

In our experiments, we used four
clusters, each with ten components, and the cluster central angles were drawn
from a uniform random distribution between $45$ degrees and $135$ degrees, where
$90$ degrees is perpendicular to the plane of the URA. The components were distributed around these central
angles according to exponential distributions with mean $7.5$ degrees, one in
each of the vertical and horizontal planes, with a $0.5$ probability of a given
component lying in either direction from the central angle in each plane. Each
component is subject to small-scale fading as $\alpha_{i \ell}$ is following a complex normal
distribution with mean $0$ and standard deviation $1$. Finally, the node
channels are normalised such that the expected power of each node's signal is
$1$.

In~\cite{el2014spatially}, clusters are generated homogeneously, however our use
case concerns channels that exhibit clear dominant directions~\cite{9539876}. To model this,
for each node, we amplified the cluster with the highest power, by multiplying
it by a scalar drawn uniformly randomly from the interval $[2.0, 6.0]$. This
could represent either the line of sight part of the node's channel, or a set of
components reflected from a highly reflective surface, if there is no
line of sight path between the base station and the node. The node channel was
then re-normalised to ensure the expected power was still $1.0$.
In this way, we generated channels with a more clear dominant component and
lower angular spectrum spread than in~\cite{el2014spatially}.

The dominant direction for each node was taken to be the central angle of the
dominant cluster. To compute the angular spectrum spread, we used the Rician
K-factor, adjusted for the angular distribution of the generated clusters for each
node. For each node $k \in \mathcal K$, the angular spectrum spread $\sigma(k)$
was computed as
\begin{equation}
    \sigma(k) = \Psi(k)\left(1.0 - \frac{P(C_d(k))}{\sum_{c \in \{\mathcal C(k) \setminus
    C_d(k)\}} P(c)} \right),
\end{equation}
where $\mathcal C(k)$ is the set of clusters in node $k$'s channel, $C_d(k) \in
\mathcal C(k)$ is the dominant cluster, $P(c)$ is the received signal power at
the base station of cluster $c \in \mathcal C(k)$, and $\Psi(k)$ is the
normalised angular spread of the clusters, computed as 
\begin{equation}\label{eq:cluster_ang_spread}
    \Psi(k) = \frac{\textrm{max}\{\psi(c) : c \in \mathcal C(k)\} -
    \textrm{min}\{\psi(c) : c \in \mathcal C(k)\}}{2\pi},
\end{equation}
where $\psi(c)$ is the central angle of cluster $c\in \mathcal C(k)$, in
radians.

We thus find the share of the total signal power in the dominant component (the
Rician K-factor), and take its complement, since a lower K-factor implies a
higher angular spectrum spread, and vice versa. We then adjust this according to how the
clusters are distributed across the angular spectrum, by multiplying by
$\Psi(k)$, the difference between the largest and smallest central angles. This
is because, as defined in Section~\ref{sec:system_model}, an angular spectrum
spread of $1.0$ should indicate a channel uniformly spread across all angles,
that is, around the entire circle, so in cases where the clusters are grouped in
a certain direction, the maximum angular spectrum spread is reduced accordingly.
Since we have considered a URA in our experiments, the
array cannot receive signals from all directions. We have used bounds of $45$ and
$135$ degrees so as to place all clusters in the region of best response for the
antenna array.

\subsection{Results}\label{sec:evaluation_results}

\begin{figure}
    \label{fig:SINR}
    \begin{subfigure}{0.5\textwidth}
    \begin{center}
	\includegraphics[width=\columnwidth]{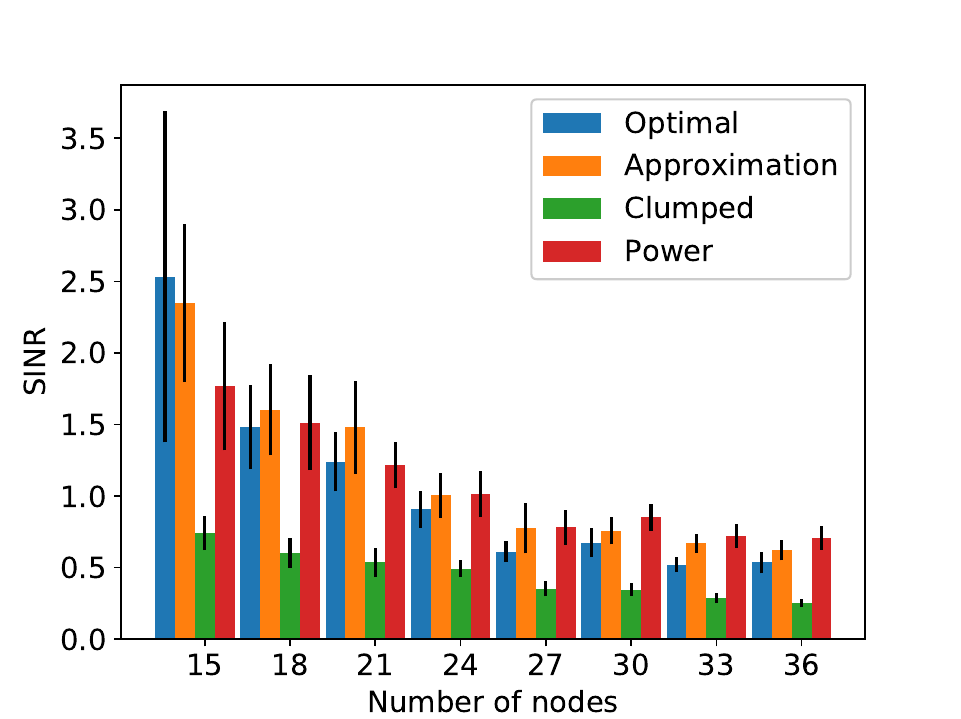}
	\caption{Minimum SINR for different numbers of nodes and node
	partitioning methods.}
	\label{fig:min_SINR}
    \end{center}
\end{subfigure}%
~
\begin{subfigure}{0.5\textwidth}
    \begin{center}
	\includegraphics[width=\columnwidth]{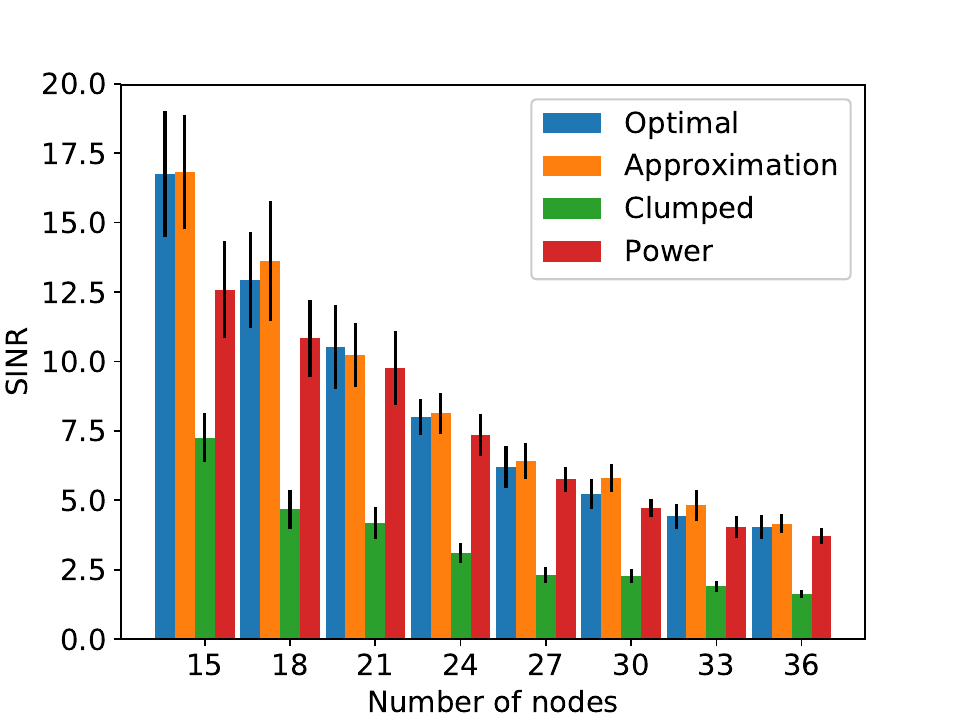}
	\caption{Average SINR for different numbers of nodes and node
	partitioning methods.}\label{fig:avg_SINR}
    \end{center}
\end{subfigure}

\vspace{0.5cm}
\begin{subfigure}{0.5\textwidth}
    \begin{center}
	\includegraphics[width=\columnwidth]{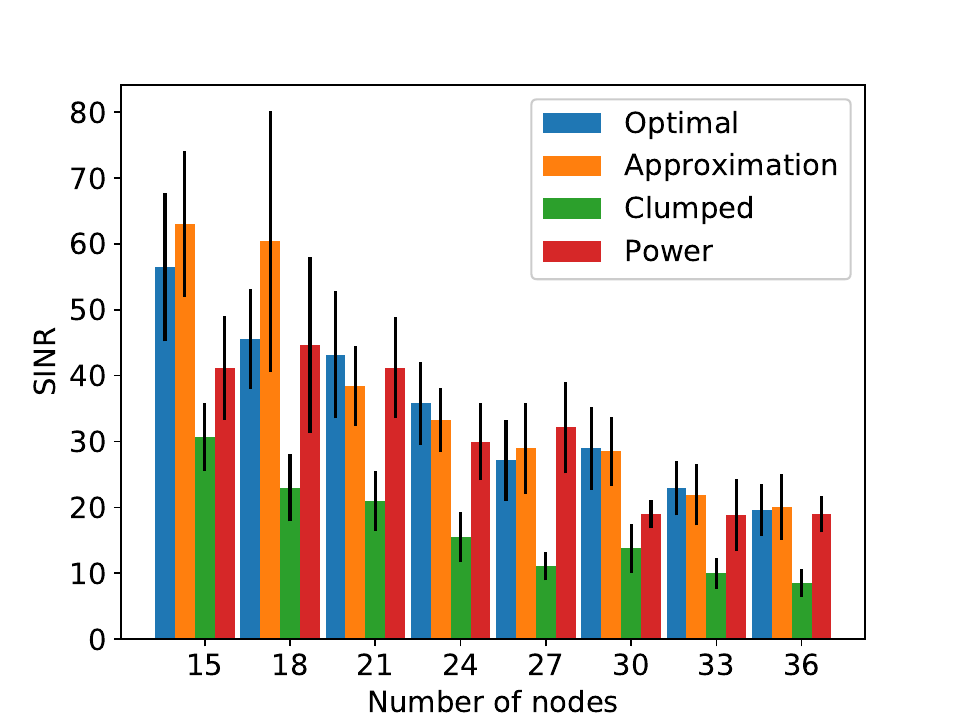}
	\caption{Maximum SINR for different numbers of nodes and node
	partitioning methods.}
	\label{fig:max_SINR}
    \end{center}
\end{subfigure}%
~
\begin{subfigure}{0.5\textwidth}
    \begin{center}
	\includegraphics[width=\columnwidth]{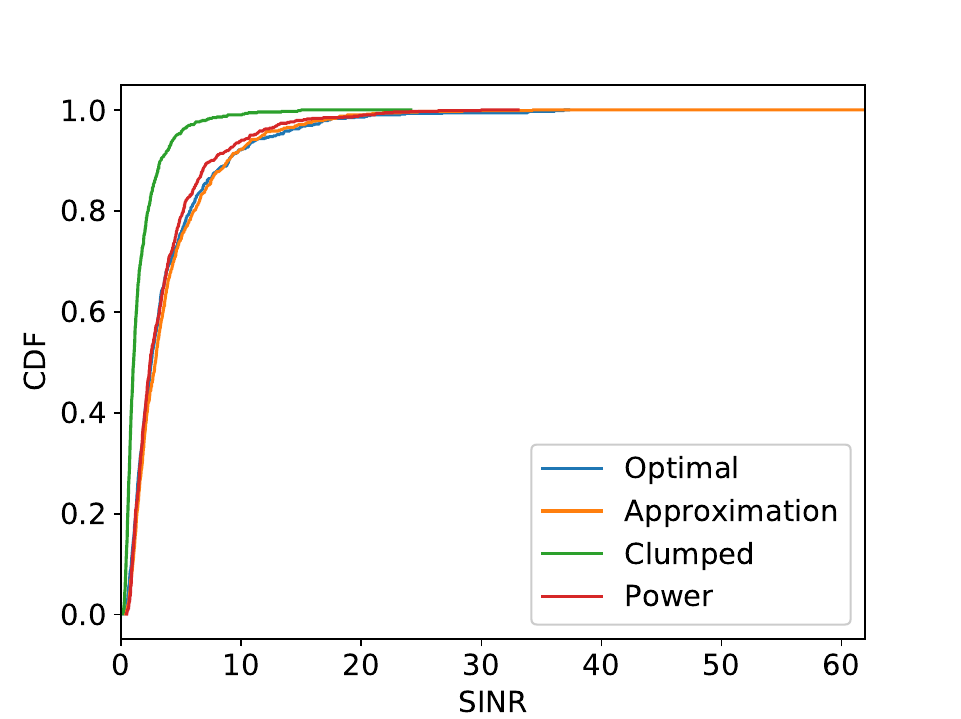}
	\caption{CDF over the SINR of all nodes for 36-node network instances.}
	\label{fig:CDF}
    \end{center}
\end{subfigure}%
    \caption{SINR performance for different numbers of nodes and partitioning
    methods.}
\end{figure}

The minimum, average, and maximum SINRs obtained for each algorithm and number
of nodes are shown in Fig.~\ref{fig:min_SINR}--\ref{fig:max_SINR}, and 
Fig.~\ref{fig:CDF} shows the CDFs of the SINRs obtained with each algorithm, 
over all nodes, in network instances with $36$ nodes. In the aggregate
figures, the values shown are the averages across all instances for each number
of nodes, with $95$\% confidence intervals. As expected, clumped partitioning
performs the worst, with the SINR decreasing as the number of nodes increases.
The results even show that a bad grouping of nodes can impact the SINR
dramatically, degrading the SINR on average by $>5$ dB for 15--20 nodes per
group in the considered scenarios. This shows the potential poor performance
that may occur if a partition is chosen that does not take into account the
directional properties of the nodes' channels. All of the other algorithms
substantially outperformed clumped partitioning on all three SINR metrics. 

The performance of the other three algorithms was similar. Even though the
optimization and approximation algorithm used only the two parameters of
dominant direction and angular spectrum spread, they were nonetheless able to match the
performance of power partitioning, which used the full channel information for
each node. This is promising, especially for IoT scenarios, as it demonstrates
that an effective node partition can be achieved even with only limited
information about the nodes' channels. Moreover, the directional properties we have
used will be relatively stable over time for static or slow-moving nodes,
meaning that the signaling needed to facilitate node partitioning is reduced,
as channel information need only be collected from the nodes occasionally. In
the best case, each time nodes transmit, the channel information obtained from
their collected pilot signals can be used to compute the next node partition,
even in cases where nodes transmit only infrequently, as in many sensor networks
and monitoring applications.

In many cases, we observe a relatively large variance in the SINRs obtained for
all algorithms. This is because while both our partitioning approaches and power
partitioning attempt to reduce inter-node interference, a true optimal partition
would require actually computing the instantaneous interference between the
nodes. This depends on the correlations between the users' channels, and while
both the directional channel properties (for our approach) and the instantaneous
power (for power partitioning) correlate with the inter-node interference, they
do not perfectly capture it. The variance is greater when the number of nodes is
lower, since with fewer nodes the probability of the nodes being more unevenly
distributed across the angular spectrum increases, leading to a higher
proportion of unusual cases.

\begin{figure}
    \label{fig:opt_perf}
    \begin{subfigure}[t]{0.45\textwidth}
    \begin{center}
	\includegraphics[width=\columnwidth]{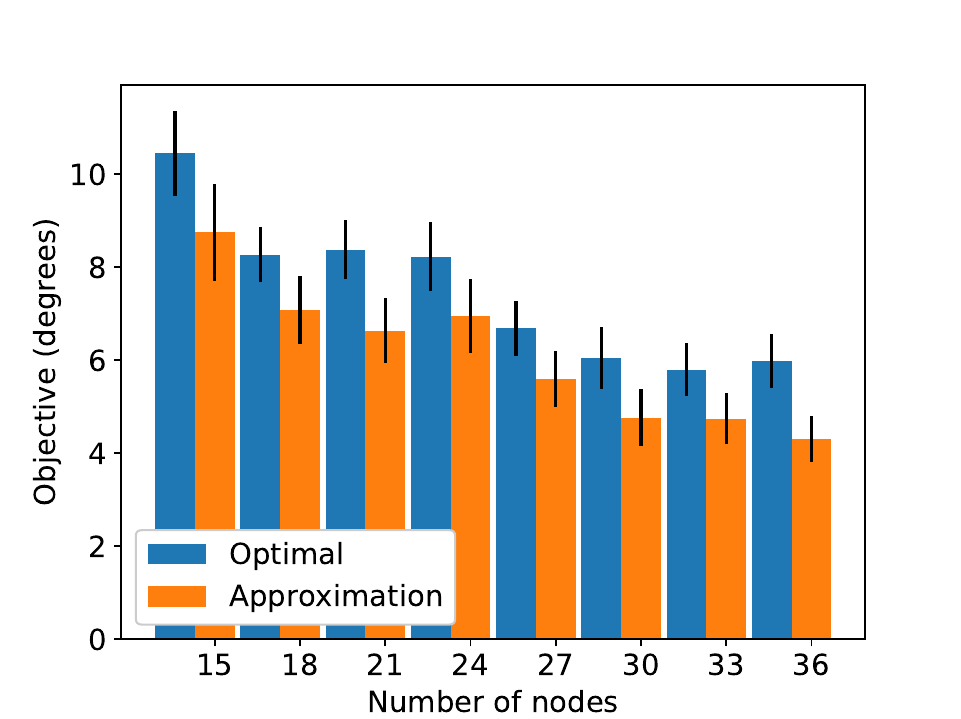}
	\caption{Objective function value for partitioning with optimization
	problem and with approximation algorithm, for different numbers of nodes.}
	\label{fig:objective}
    \end{center}
\end{subfigure}%
~
\begin{subfigure}[t]{0.45\textwidth}
    \begin{center}
	\includegraphics[width=\columnwidth]{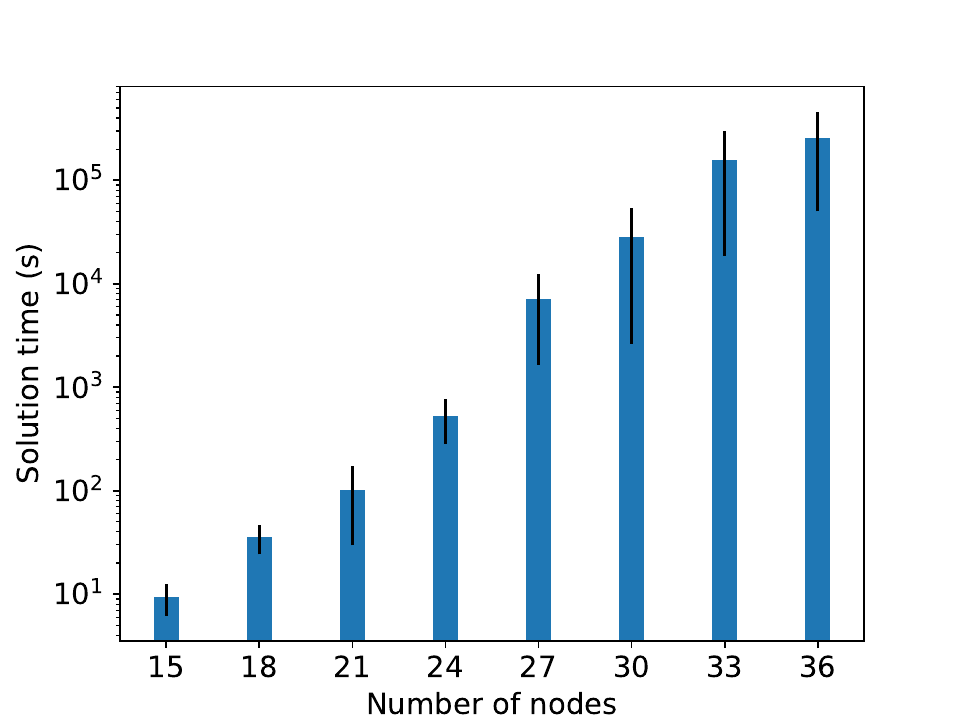}
	\caption{Solution time vs. number of nodes.}
	\label{fig:time_num_nodes}
    \end{center}
\end{subfigure}
    \caption{Performance of the optimization and approximation algorithm.}
\end{figure}

From the figures, we can see that the optimization and approximation algorithm
performed similarly, with the approximation even obtaining a higher SINR in some
cases, although the difference between the two is only statistically significant
in one case, for the minimum SINR for $33$ nodes. The reason that the
approximation algorithm can achieve better performance than the optimization is
again that here we measure performance in terms of SINR, calculated from the
complete node channels, whereas the two algorithms --- optimization and
approximation --- have only reduced channel information available to them in the
form of the dominant directions and angular spectrum spreads. In other words, 
the objective function the two algorithms aim for is not the same as the SINR
measures plotted in the figure, only a reasonable proxy for it with reduced
channel information. As can be seen in Fig.~\ref{fig:objective}, in terms of the
objective function value obtained, the optimization always outperforms the
approximation, as expected. Across all the cases tested, the approximation
algorithm achieves an average of $81$\% of the optimal objective function value.

Fig.~\ref{fig:time_num_nodes} shows the time to solve
formulation~\eqref{form:pizza} to optimality versus the number of nodes. As is
typical for this kind of combinatorial optimization problem, the solution time
grows exponentially with the problem instance size (note that the solution times
are shown on a log scale). For small instances of $15$ nodes, the optimization
takes less than ten seconds to run, which is feasible in practice for IoT
scenarios where nodes transmit on the order of minutes or higher. However, for
larger instances, it would not be feasible to use the optimization; the largest
solution time observed was $1232518.65$ s --- more than $14$ days ---- and
realistic IoT scenarios can have a far greater number of nodes than we tested
here. As a comparison, our Python implementation of the approximation algorithm
ran in less than $0.1$ ms for all cases tested. (The other comparison algorithms had similarly fast solution times as well.) For practical deployments, the
approximation algorithm is thus a more appropriate solution. Based on the SINR characteristics for the grouped nodes, further communication-oriented performance paramaters such as outage probability and achievable rate can be derived for state-of-the-art transmission schemes~\cite{marzetta2016fundamentals}.

\section{Conclusion and Future Work}\label{sec:conclusion}

In this paper we have developed a new method for node grouping in massive MIMO
systems based on directional channel characteristics. The two channel properties
we use, dominant direction and angular spectrum spread, are typically stable
over large time scales, when compared to the full channel information that
should be updated at the rate of the coherence time of the channel. Our approach
hence enables grouping of low-power IoT devices with minimal signalling
overhead, even when the devices' measurement and transmission periods are long.
We provided both a mixed-integer optimization formulation and an efficient
approximation algorithm to perform the node grouping. In our numerical
evaluation, we demonstrated that our approach provides good performance, in
terms of the minimum SINR achieved for any node in any group, comparable to a
reference method exploiting full CSI. Our assessment also
highlights that a partitioning that groups nodes with similar directional
properties together can have a particularly detrimental effect on link quality due
to the interference generated by MRC precoding.

A light signalling user grouping approach as we have demonstrated in this paper
is valuable for any situation where signalling overhead is significant. This
will be the case when there are a very large number of nodes, as in IoT
scenarios. At the end devices, reducing signalling overhead also reduces energy
usage, which is critical for battery powered IoT devices. Although we have thus far only tested cases with up to 36 nodes, and in real IoT deployments the number of nodes is expected to be much greater, the approximation algorithm we have presented shows good scalability and should thus be suitable even in larger scenarios.

In future work, we aim to close the loop by performing measurements in
which we measure nodes' channels and then apply our solutions and
measure the performance of the resulting node groups. We also plan to extend
our method to take into account errors in the directional channel information. This
will not only make the method more robust in case of measurement errors, but
also cater to mobile nodes, since as a node moves, its directional information
changes, thus effectively producing an error compared to the last time this
information was measured. 

The range of scenarios investigated can also be expanded in future work. By focusing on heuristic methods, larger problems instances can be tested, as well as more frequency bands and different numbers of antennas at the base station. Multi-cell and distributed, cell-free systems are also an interesting direction in future, in particular the latter as cell-free architectures are likely to feature prominently in 6G systems, and this further complicates the node grouping problem, since nodes cannot just be grouped within each cell.

\balance

\reftitle{References}
\externalbibliography{yes}
\bibliography{references}

\end{document}